\documentclass[twocolumn,amsmath,amssymb,floatfix,nofootinbib]{revtex4}

\usepackage{epsfig}
\usepackage{amsmath,amssymb}
\usepackage[all]{xy}

\newcommand{\beq}{\begin{equation}}
\newcommand{\eeq}{\end{equation}}
\newcommand{\lsi}{\,\raisebox{-0.13cm}{$\stackrel{\textstyle<}
{\textstyle\sim}$}\,}
\newcommand{\gsi}{\,\raisebox{-0.13cm}{$\stackrel{\textstyle>}
{\textstyle\sim}$}\,}
\newcommand{\be}{\begin{equation}}
\newcommand{\ee}{\end{equation}}

\begin{document}

\title{\bf Transitions of two baryons to the H dibaryon in nuclei}
\author{Glennys R. Farrar and Gabrijela Zaharijas}
\affiliation{\it Center for Cosmology and Particle Physics\\
New York University, NY, NY 10003,USA}

\begin{abstract}
We calculate the suppression in the rate at which two baryons in
a nucleus (viz., nucleons or $\Lambda$'s) convert to an H
dibaryon, using an Isgur-Karl wavefunction for quarks in the
baryons and H, and a Bethe-Goldstone wavefunction for the
baryons in the nucleus.  If $r_H \lsi 1/3~ r_N$, we find
$\tau_{A_{\Lambda\Lambda}\rightarrow A'_H}\gsi \tau_\Lambda$ and
the observation of $\Lambda$ decays from double-$\Lambda$
hypernuclei does not exclude the existence of the H.  If $m_H <
2 m_p$, nuclei are unstable but have very long lifetimes.  For
reasonable values of $r_H$ and the nuclear wavefunction, the
lifetime can be long enough to evade anticipated SuperK limits
$\tau_{A_{NN}\rightarrow A'_H}\gsi {\rm few} 10^{29}$ yr, or
short enough to be observed. An analysis of SuperK data to look
for this possibility should be undertaken.
\end{abstract}

 \maketitle

%\vspace{4pt}
%\vfill\eject

\section{Introduction}

The H dibaryon corresponds to the most symmetric color-spin
representation of six quarks ($uuddss$). It is a flavor singlet
state with charge 0, strangeness -2 and spin-isospin-parity
$I(J^P)=0(0^+)$. The existence of the  H was predicted by Jaffe
in 1977 \cite{jaffe} in the framework of the quark-bag model.
Its mass was originally estimated to be around 2150 MeV, making
it stable toward strong decay to two $\Lambda$ particles. Since
then, there have been many theoretical efforts to determine its
mass and production cross section and, on the experimental side,
many inconclusive or unsuccessful attempts to produce and detect
it.

Our work is prompted by the possibility that the H is lighter
than two nucleons. This is motivated by several lines of
reasoning based on hadron phenomenology and non-perturbative QCD
modeling\cite{f:stableH}.  The interpretation of $\Lambda(1405)$
and $\Lambda(1520)$ as bound states of gluon plus $uds$ quarks
in a flavor-singlet color-octet state, suggests the H has
properties similar to a glueball and $m_H \approx 1.3-1.8 $
GeV\cite{kf:Lam1405,f:stableH}.  Adding current quark masses to
the original Skyrme model calculation with massless
quarks\cite{balachandran:skyrmeH,f:stableH} gives $m_H \approx
1.8$ GeV.  Finally, an instanton-liquid calculation gives $m_H =
1780$ MeV\cite{kochelev:H}.

Being tightly bound, the H is expected to be a spatially compact
state. Analogy with the glueball and instanton-liquid results
suggest $r_H \approx r_G \approx (1/3-1/2)~ r_\pi \approx (1/6-1/4)~r_N $ \cite{f:stableH,shuryak:RG}.  In the absence of an
unquenched, high-resolution lattice QCD calculation capable of a
reliable determination of the H mass and size, we will take
$r_H/r_N = 1/f$ with $f$ treated as a parameter expected to be
in the range 4-6.  For a more detailed discussion of motivation
and properties of a stable H, and a review of experimental
constraints on such an H, see ref. \cite{f:stableH}.

In this paper we focus on two types of experimental constraints
on the transition of two baryons to an H in a nucleus,
$A_{BB}\rightarrow A'_{H}X$. Experiments observing single
$\Lambda$ decays from double $\Lambda$ hypernuclei $A_{\Lambda
\Lambda}$\cite{ags,kek2} indicate that $\tau(A_{\Lambda
\Lambda}\rightarrow A_{H}'X) \lsi \tau_\Lambda = 3 ~ 10^{-10}$
sec.  In addition, if the H is lighter than two nucleons, nuclei
are unstable toward $\Delta S=-2$ weak decays producing the H
particle in the final state.  To estimate the rates for these
processes requires calculating the overlap of initial and final
quark wavefunctions in a nucleus.  We will model that overlap
using an Isgur-Karl harmonic oscillator model for the baryons
and H, and the Bethe-Goldstone wavefunction for a nucleus.  Our
results will be expressed in terms of the parameter $f=r_N/r_H$
and the nuclear hard core radius.  We will find that the
stability of nuclei is the more stringent constraint on the
properties of a stable H, but is acceptably small if the H is
sufficiently compact: $r_H \lsi ~1/4 ~r_N$ depending on mass and
nuclear hard core radius. Adequate suppression of
$\Gamma(A_{\Lambda \Lambda}\rightarrow A_{H}'X)$ requires $r_H
\lsi 1/3 ~r_N$, whether H is stable or not.  Thus a byproduct of
this investigation is that the conventional H with mass $2 m_N <
m_H < 2 m_\Lambda$ may still be viable in spite of the
observation of double-$\Lambda$ hypernuclei, consistent with the
conclusion of ref. \cite{kahana}.

This paper is organized as follows. In section \ref{expts} we
describe in greater detail the two types of experimental
constraints on the conversion of baryons to an H in a nucleus.
In section \ref{rates} we calculate lifetimes for these
transitions, using a non-relativistic harmonic oscillator quark
model and a phenomenological treatment of the weak interactions.
The results are summarized in section \ref{summary}.

\section{Experimental constraints}
\label{expts}
\subsection{Stability of nuclei}

There are a number or possible reactions by which two nucleons
can convert to an H in a nucleus.  The initial state is most
likely to be $pn$ or $nn$ in a relative s-wave, because in other
cases the Coulomb barrier or relative orbital angular momentum
suppresses the overlap of the nucleons at short distances which
is necessary to produce the H. If $m_H \lsi 2 m_N - n
m_\pi$\footnote{Throughout, we use this shorthand for the more
precise inequality $m_H < m_{A} - m_{A'} - m_X$ where $m_X$ is
the minimum invariant mass of the final decay products.}, the
final state can be $H \pi^+ $ or $H \pi^0$ and $n-1$ pions with
total charge 0.  For $m_H \gsi 1740$ MeV, the most important
reactions are $p n \rightarrow H e^+ \nu_e$ or the
radiative-doubly-weak reaction $n n \rightarrow H \gamma$.

The most sensitive experiments to place a limit on the stability
of nuclei are proton decay experiments. Super Kamiokande
(SuperK), places the most stringent constraint due to its large
mass;  it is a water Cerenkov detector with a 22.5 kiloton
fiducial mass, corresponding to $8~10^{32}$ oxygen nuclei.
SuperK is sensitive to proton decay events in over 40 specific
proton decay channels\cite{SuperK}. Since the signatures for the
transition of two nucleons to the H are substantially different
from the monitored transitions, a specific analysis by SuperK is
needed to place a limit.  We will discuss the order-of-magnitude
of the limits which can be anticipated.

Detection is easiest if the H is light enough to be produced
with a $\pi^+$ or $\pi^0$. The efficiency of SuperK to detect
neutral pions, in the energy range of interest (KE = 0-$\sim
300$ MeV), is around 70 percent. In the case that a $\pi ^+$ is
emitted, it can charge exchange to $\pi ^0$ within the detector,
or be directly detected as a non-showering muon-like particle
with similar efficiency.  More difficult is the most interesting
mass range $m_H \gsi 1740$ MeV, for which the dominant channel
$p n \rightarrow H e^+ \nu$ gives an electron with $E \sim (2
m_N - m_H)/2 \lsi  70$ MeV.  With a rate of order $\alpha$
smaller, the $nn \rightarrow H \gamma$ channel would give a
monochromatic photon with energy $(2 m_N - m_H) \lsi 100$ MeV.

We can estimate SuperK's probable sensitivity as follows.  The
ultimate background comes primarily from atmospheric neutrino
interactions: $\nu N \rightarrow N'(e,\mu),\quad \nu N
\rightarrow N'(e,\mu)+n\pi ,\quad \nu N\rightarrow \nu N' +n\pi
$, which has a rate of about 100 ${\rm kton} ^{-1} {\rm
yr}^{-1}$.  Without a strikingly distinct signature, it would be
difficult to detect a signal rate significantly smaller than
this, which would imply SuperK might be able to achieve a
sensitivity of order $\tau_{A_{NN}\rightarrow A_{H}'X}\gsi {\rm
few} 10^{29}$ yr. Since the H production signature is not more
favorable than the signatures for proton decay, the SuperK limit
on $\tau_{A_{NN}\rightarrow A_{H}'X}$ can at best be $0.1
\tau_p$, where $0.1$ is the ratio of Oxygen nuclei to protons in
water. Detailed study of the spectrum of the background is
needed to make a more precise statement. We can get a lower
limit on the SuperK lifetime limit by noting that the SuperK
trigger rate is a few Hz\cite{SuperK}, putting an immediate
limit $\tau_{O\rightarrow H + X }\gsi {\rm few} 10^{25}$ yr,
assuming the decays trigger SuperK.

While SuperK limits depend on specific decay channels, three
other experiments potentially establish limits on the proton
lifetime which are independent of the decay channel\cite{pdb}.
They place weaker constraints on the lifetime, due to their
smaller size, but are of interest because they measure the
stability of nuclei directly.  The experiments of Dix et
al.\cite{dix} and Evans et al.\cite{evans} are not sensitive to
two nucleon transitions and thus are not applicable to nuclei
disintegrating with the emission of an H.

The experiment of Flerov et. al.\cite{flerov} could in principle
be sensitive to such transitions.  It searched for decay
products from ${\rm Th}^{232}$, above the Th natural decay mode
background of 4.7 MeV $\alpha$ particles, emitted with the rate
$\Gamma _{\alpha}=0.7~10^{-10} {\rm
yr}^{-1}$. The conversion of two nucleons in ${\rm Th}^{232}$ could result in the following decay chains:\\
${\rm Th}^{232} _{pp}\rightarrow {\rm Ra}^{230}_H X;
 $$\xymatrix{{\rm Ra}^{230} \ar[r]^{93 min} &{\rm
Ac}^{230}}+\beta (0.99 MeV)$
${\rm Th}^{232} _{pn}\rightarrow
{\rm Ac}^{230}_H X; $$\xymatrix{{\rm Ac}^{230} \ar[r]^{122
s}&{\rm Th}^{230}}+\beta (2.7 MeV)$
${\rm Th}^{232} _{nn}\rightarrow  {\rm Th}^{230}_H X;$$\xymatrix{{\rm Th}^{230} \ar[r]^{75380yr}& {\rm Ra}^{226}}+\alpha (4.7 MeV).$\\
However in general the transitional nuclear state, denoted e.g.,
${\rm Ra}^{230}_H$, would have additional more complicated decay
chains through excited states.  Note that the H does not bind to
nuclei\cite{fz:nucbind}; it simply recoils with some momentum
imparted in its production. The Flerov et al\cite{flerov}
experiment must have cuts to remove the severe background of 4.7
MeV $\alpha$'s.  If these cuts do not remove the events with
production of an H, it would imply the limit $\tau_{{\rm
Th}^{232}\rightarrow H + X}> 10^{21}$ yr.  Unfortunately ref.
\cite{flerov} does not discuss these cuts or the experimental
sensitivity in detail.  An attempt to correspond with the
experimental group, to determine whether their results are
applicable to the H, was unsuccessful.

\subsection{Double  $\Lambda $ hyper-nuclei detection}

There are five experiments which have reported positive results
in the search for single $\Lambda$ decays from double $\Lambda$
hypernuclei. We will describe them briefly. The three early
emulsion based experiments \cite{prowse,danysz,kek} suffer from
ambiguities in the particle identification, and therefore are
considered less reliable. In the latest emulsion experiment at
KEK \cite{kek2}, a double hypernucleus event has been observed
and interpreted with good confidence as the sequential decay of
${\rm He}^6 _{\Lambda \Lambda}$ emitted from a $\Xi ^-$ hyperon
nuclear capture at rest. The binding energy of the double
$\Lambda$ system is obtained in this experiment to be
$B_{\Lambda \Lambda }=1.01\pm 0.2$ MeV, in significant
disagreement with the results of previous emulsion experiments,
finding $B_{\Lambda \Lambda }\sim 4.5$ MeV.

The synchrotron based experiment \cite{ags} used the $(K^-,
K^+)$ reaction on a ${\rm Be}^9$ target to produce S=-2 nuclei.
That experiment detected pion pairs, coming from the same vertex
in the Be target. Each pion in a pair indicates one unit of
strangeness change from the (presumably) di-$\Lambda$ system.
Peaks in the two pion spectrum have been observed, interpreted
as corresponding to two kinds of decay events. The pion kinetic
energies in those peaks are (114,133) MeV and (104,114) MeV. The
first peak can be understood as two independent single $\Lambda$
decays from $\Lambda \Lambda$ nuclei. The energies of the second
peak do not correspond to known single $\Lambda$ decay energies
in hyper-nuclei of interest. The proposed explanation\cite{ags}
is that they are pions from the decay of the double $\Lambda$
system, through some specific He resonance. The required
resonance has not yet been observed experimentally, but its
existence is considered plausible. This experiment does not
suffer from low statistics or inherent ambiguities, and one of
the measured peaks in the two pion spectrum suggests observation
of consecutive weak decays of a double $\Lambda$ hyper-nucleus.
The binding energy of the double $\Lambda$ system $B_{\Lambda
\Lambda }$ could not be determined in this experiment.

The KEK and BNL experiments demonstrate quite conclusively, in
two different techniques, the observation of $\Lambda$ decays
from double $\Lambda$ hypernuclei.  Therefore $\tau _{A_{\Lambda
\Lambda}\rightarrow A_{H}'X}$ cannot be much less than $\approx
10^{-10}$s.  (To give a more precise limit on $\tau _{A_{\Lambda
\Lambda}\rightarrow A_{H}'X}$ requires a detailed analysis by
the experimental teams, taking into account the number of
hypernuclei produced, the number of observed $\Lambda$ decays,
the acceptance, and so on.)  As will be seen below, this
constraint is readily satisfied if the H is compact: $r_H \lsi
1/3 ~r_N$.

\section{BB to H transition rates in nuclei} \label{rates}

As discussed in the introduction, the H may be considerably more
compact than a nucleon which would suppress its production from
a two baryon initial state, due to the small overlap of the
initial and final wave functions in position space. We will
estimate this suppression using the non-relativistic harmonic
oscillator quark model. Additional suppression comes from the
hard core repulsion in the nucleon-nucleon wave function in the
nucleus. To take that into account, we use the Bethe-Goldstone
relative wave function of two baryons in a nucleus.

The matrix element for the transition $A_{NN} \rightarrow A'_H X
$ is calculated in the $\Lambda \Lambda$ pole approximation, as
a product of matrix elements for two subprocesses: a transition
matrix element for formation of the H from a $\Lambda \Lambda$
system, $ |{\cal M}| _{\Lambda \Lambda \rightarrow H~X}$, times
the amplitude for a weak doubly-strangeness-changing transition,
$|{\cal M}|_{NN \rightarrow \Lambda \Lambda}$.  The suppression
in the spatial wavefunction overlap enters the $\Lambda \Lambda
\rightarrow H$ transition. We calculate this part of the
transition amplitude first. The estimate of $|{\cal M}|_{NN
\rightarrow \Lambda \Lambda X }$ based on weak interaction
phenomenology is given afterwords.

\subsection{Calculation of $|{\cal M}| _{\Lambda \Lambda \rightarrow H}$}

We calculate $|{\cal M}| _{\Lambda \Lambda \rightarrow H}$ in
position space as the overlap of the H and $\Lambda \Lambda$
wave functions in the Isgur-Karl (IK) non-relativistic harmonic
oscillator quark model\cite{faiman,bhaduri}.  We take the
$\Lambda$ spatial wavefunction to be the same as the nucleon's.
For now we are concerned with the dynamics of the process and we
defer discussion of the suppression from the spin-flavor part of
the transition amplitude.

The IK model was designed to reproduce the masses of the
observed resonances and it has proved to be successful in
calculating baryon decay rates \cite{faiman}. In the IK model,
the quarks in a baryon are described by the Hamiltonian \beq
\label{hamiltonian} H=\frac {1}{2m} (p^2 _1+p^2 _2+p^2 _3)
+\frac{1}{2}K\Sigma_{i<j} ^3 (\vec {r}_i -\vec {r}_j)^2 \eeq
where we have neglected constituent quark mass differences.  The
wave function of
 baryons can then be written in terms of the relative positions of
quarks, while the center of mass motion is factored out. The
relative wave function in this model is\cite{faiman,bhaduri}
\beq \Psi _{B} (\vec{r}_1,\vec{r}_2,\vec{r}_3) = N_{B} \exp
\left[ {-\frac {\alpha_{B} ^2}{6}\Sigma_{i<j} ^3 (\vec {r}_i
-\vec {r}_j)^2}\right] \eeq where $N_B$ is the normalization
factor, $\alpha _B=\frac {1}{\sqrt{<r_B ^2>}}=\sqrt{3Km}$, and
$<r_B ^2>$ is the baryon mean charge radius squared. Changing
variables to \beq \label{rholambda} \vec {\rho} =\frac {\vec
{r_1} -\vec {r_2}}{\sqrt{2}},~\vec {\lambda}=\frac {\vec {r_1}
+\vec {r_2}-2 \vec {r_3}}{\sqrt{6}} \eeq reduces the wave
function to two independent harmonic oscillators. In the ground
state \beq \Psi_{B} (\vec {\rho}, \vec {\lambda})=\left( \frac
{\alpha_B}{\sqrt{\pi}} \right) ^3 \exp\left[ -\frac
{\alpha_{B}^2}{2} (\rho ^2 + \lambda ^2)\right]. \eeq

One of the deficiencies of the IK model is that the value of the
$\alpha _B$ parameter needed to reproduce the mass splittings of
lowest lying $\frac {1}{2} ^+$ and $\frac {3}{2} ^+$ baryons
corresponds to a mean charge radius squared for the proton of
$<r^2_{ch}>= \frac {1}{\alpha _B ^2}=0.49$ fm. This is
distinctly smaller than the experimental value of 0.86 fm. Our
results depend strongly on the choice of $\alpha_B$ and
therefore we should keep in mind this problem.  Another concern
is the applicability of the non-relativistic IK model in
describing quark systems, especially in the case of the tightly
bound H. With $r_H/r_N = 1/f$, the quark momenta in the H are
$\approx f$ times higher than in the nucleon, which makes the
non-relativistic approach more questionable than in the case of
nucleons. Nevertheless we adopt the IK model because it offers a
tractable way of obtaining a qualitative estimate of the effect
of the small size of the H on the transition rate, and there is
no other alternative available at this time.

We fix the wave function for the H particle starting from the
same Hamiltonian (\ref{hamiltonian}), but generalized to a six
quark system.  For the relative motion part this gives \beq
\Psi_{H}=N_{H}\exp\left[-\frac{\alpha_{H}^2}{6}\sum _{i<j} ^6
(\vec{r_i} -\vec{r_j})^2\right]. \eeq The space part of the
matrix element $<A'_{H}|A_{ \Lambda \Lambda }>$ is given by the
integral \beq \int \prod _{i=1} ^6 d^3\vec{r}_i \Psi _{\Lambda}
^{a} (1,2,3) \Psi _{\Lambda} ^{b} (4,5,6) \Psi_H (1,2,3,4,5,6).
\eeq We can rewrite this in a more convenient form, changing
variables to \beq
\vec{r}_1,\vec{r}_2,\vec{r}_3,\vec{r}_4,\vec{r}_5,\vec{r}_6
\rightarrow \vec
{\rho}^{a},\vec{\lambda}^{a},\vec{\rho}^{b},\vec {\lambda}^{b},
\vec {a}, \vec {R}_{CM} \eeq where $\vec {\rho}^{a(b)}$ and
$\vec {\lambda}^{a(b)}$ are defined as in eq (\ref{rholambda}),
with $a(b)$ referring to coordinates $1,2,3~(4,5,6)$.  (Since we
are ignoring the flavor-spin part of the wavefunction, we can
consider the six quarks as distinguishable and not worry about
fermi statistics at this stage.)  We also define the
center-of-mass position and the separation, $\vec {a}$, between
initial baryons $a$ and $b$: \beq \vec {R}_{CM}=\frac {\vec
{R}_{CM}^{a}+\vec {R}_{CM}^{b}}{2},~ \vec {a}=\vec
{R}_{CM}^{a}-\vec {R}_{CM}^{b}. \eeq Using these variables, the
H ground state wave function becomes
\begin{eqnarray}
\Psi_{H}&=&\left( \frac{3}{2^3}\right) ^{1/4} \left( \frac{\alpha _H}{\sqrt{\pi}} \right)^{15/2}\\
&\times & \exp[-\frac {\alpha_{H} ^2}{2} (\vec {\rho^{a}}^2
+ \vec {\lambda ^{a}}^2+\vec {\rho^{b}}^2 + \vec {\lambda
^{b}}^2 +\frac {3}{2} \vec {a}^2)] \nonumber
\end{eqnarray}
and the overlap of the space wave functions is given by \beq
|{\cal M}|_{\Lambda \Lambda \rightarrow H}=\int \prod _{i=a,b}
d^3 \rho^i d^3 \lambda ^i d^3 a~ \psi _H \psi ^a
_{\Lambda}~\psi^b _{\Lambda}~ \psi _{nuc} \eeq where the center
of mass dependence has been factored out, $\psi ^{a,b}
_{\Lambda}=\psi ^{a,b} _{\Lambda}(\vec {\rho}^{a,b},\vec
{\lambda}^{a,b})$, and $\psi _{nuc}= \psi _{nuc} (\vec {a})$ is
the relative wavefunction function of the two $\Lambda 's$ in
the nucleus. The integration over the center of mass position of
the system gives a 3 dimensional momentum delta function.  In
the case of pion or lepton emission, plane waves of the emitted
particles should be included in the integrand. For brevity we
use here the zero momentum transfer, $\vec {k} =0$
approximation, which we have checked holds with good accuracy;
this is not surprising since typical momenta are $\lsi 0.3$ GeV.

To describe two $\Lambda$'s or nucleons in a nucleus we will use
solutions of the Bruecker-Bethe-Goldston equation describing the
interaction of a pair of fermions in an independent pair
approximation; see, e.g., \cite{walecka}.  The two particle
potential in a nucleus is poorly known at short distances.
Measurements (the observed deuteron form factors, the sums of
longitudinal response of light nuclei,...) only constrain the
two-nucleon potentials and the wave functions they predict at
internucleon distances larger than $0.7$ fm
\cite{pandharipande}.  The Bethe-Goldstone equation can be
solved analytically when a hard-core potential is used.  While
the hard-core form is surely only approximate, it is useful for
our purposes because it enables us to isolate the sensitivity of
the results to the short-distance behavior of the wavefunction.
We stress again, that more ``realistic" wavefunctions are in
fact not experimentally constrained for distances below 0.7 fm.
Rather, their form at short distance is chosen for technical
convenience or aesthetics.

For the s-wave, the B-G wavefunction is
\beq
\Psi_{BG}(\vec{a})=\left\{
\begin{array}{ll}
N_{BG}\frac{u(k_F a)}{k_F a} & \textrm{for \quad $a>\frac{c}{k_F}$} \\
0  & \textrm {for $\quad a<\frac{c}{k_F}$}
\end{array}\right.
\eeq where $\frac{c}{k_F}$ is the hard core radius. The function
$u$ vanishes at the hard core surface by construction. It then
rapidly approaches the unperturbed value $1$, crossing over that
value at the so called ``healing distance''.  Expressions for
$u$ and $N_{BG}$ can be found in \cite{walecka}.

After performing the Gaussian integrals analytically, the
overlap of the space wave functions becomes
\begin{eqnarray}
|{\cal M}|_{\Lambda \Lambda \rightarrow H}&=&\frac {1}{4} \left (\frac
{2f}{1+f^2}\right )^6 \left( \frac{3}{2^3} \right)^{1/4}\left( \frac{\alpha _H}{\sqrt{\pi}} \right)^{3/2}\\ \nonumber
&\times & N_{BG}\int^{\infty} _{\frac{c}{k_F}} d^3 a
\frac {u(k_F a)}{k_F a}e ^{-\frac {3}{4}\alpha_{H} ^2 a^2}
\end{eqnarray}
where the factor 1/4 comes from the probability that two
nucleons are in a relative s-wave, and $f$ is the
previously-introduced ratio of nucleon to H radius; $\alpha
_H=f~\alpha _B $. Since $N_{BG}$ has dimensions  $V^{-1/2}$ the
spatial overlap $|{\cal M}| _{\Lambda \Lambda \rightarrow H}$ is
a dimensionless quantity, characterized by the ratio $f$, the
Isgur-Karl oscillator parameter $\alpha_B$, and the value of the
hard core radius.  It is shown in Fig. 1 for a range values of
hard-core radius and $f$, using the standard value of $\alpha_B$
for the IK model\cite{bhaduri}.

\vspace{5mm}
\centerline{\epsfig{file=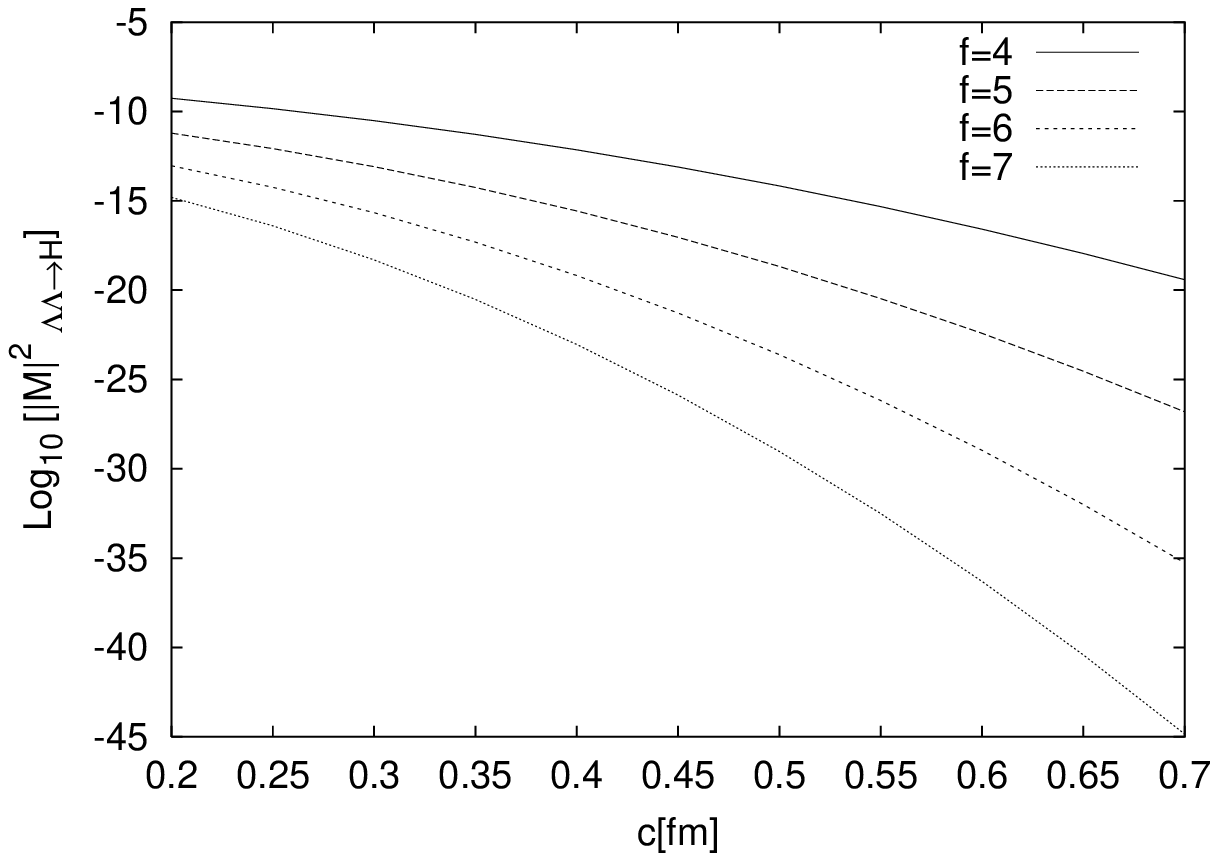,width=8cm}}{\footnotesize\textbf{Figure
1:} Log$_{10}$ of $|{\cal M}|^2_{\Lambda \Lambda \rightarrow H}$
versus hard core radius in fm, for ratio $f=R_N/R_H=$ 4, 5, 6,
7.}
\vspace{5mm}\\

\subsection{Lifetime of doubly-strange nuclei}

We can now estimate the decay rate of a doubly-strange nucleus:
\begin{eqnarray}
\Gamma_{A_{\Lambda \Lambda} \rightarrow A'_{H} \pi} &\approx&
K^2(2\pi )^4 \frac {m^2 _q }{2(2m_{\Lambda \Lambda})} \\
\nonumber &\times& \Phi _2 |{\cal M}|^2_{\Lambda \Lambda
\rightarrow H}.
\end{eqnarray}
where $\Phi _2$ is the two body phase final space factor,
defined as in \cite{pdb}, and $m_{\Lambda \Lambda}$ is the
invariant mass of the $\Lambda$'s, $\approx 2 m_{\Lambda}$. The
factor $K$ contains the transition element in spin flavor space.
It can be estimated by counting the total number of flavor-spin
states a $uuddss$ system can occupy, and taking $K^2$ to be the
fraction of those states which has the correct quantum numbers
to form the H. That gives $K^2\sim 1/1440$. Thus we obtain the
lifetime estimate \beq \tau_{A_{\Lambda \Lambda}\rightarrow
A'_{H} \pi}\approx \frac {1.4}{K^2 |{\cal M}|^2_{\Lambda \Lambda
\rightarrow H}}10^{-21}~ {\rm s}, \eeq where the phase space
factor was calculated for $m_H = 1.5$ GeV.

Fig. 2 shows $|{\cal M}|^2_{\Lambda \Lambda \rightarrow H}$ in
the range of $f$ and hard-core radius where its value is in the
neighborhood of the experimental limits. Evidently, $|{\cal
M}|^2_{\Lambda \Lambda \rightarrow H} \lsi 10^{-8}$ is satisfied
even for relatively large H, e.g., $r_H \lsi 1/3 ~r_N $ for the
canonical choice 0.4 fm for hard-core radius. This suppresses
$\Gamma(A_{\Lambda \Lambda}\rightarrow A'_{H} X)$ sufficiently
that some $\Lambda$'s in a double-$\Lambda$ hypernucleus will
decay prior to formation of an H. Thus the observation of single
$\Lambda$ decay products from double-$\Lambda$ hypernuclei
cannot be taken to exclude the existence of an H with mass below
$2 m_\Lambda$ unless it can be demonstrated that $r_H \geq
1/3~r_N$.

\vspace{5mm}
\centerline{\epsfig{file=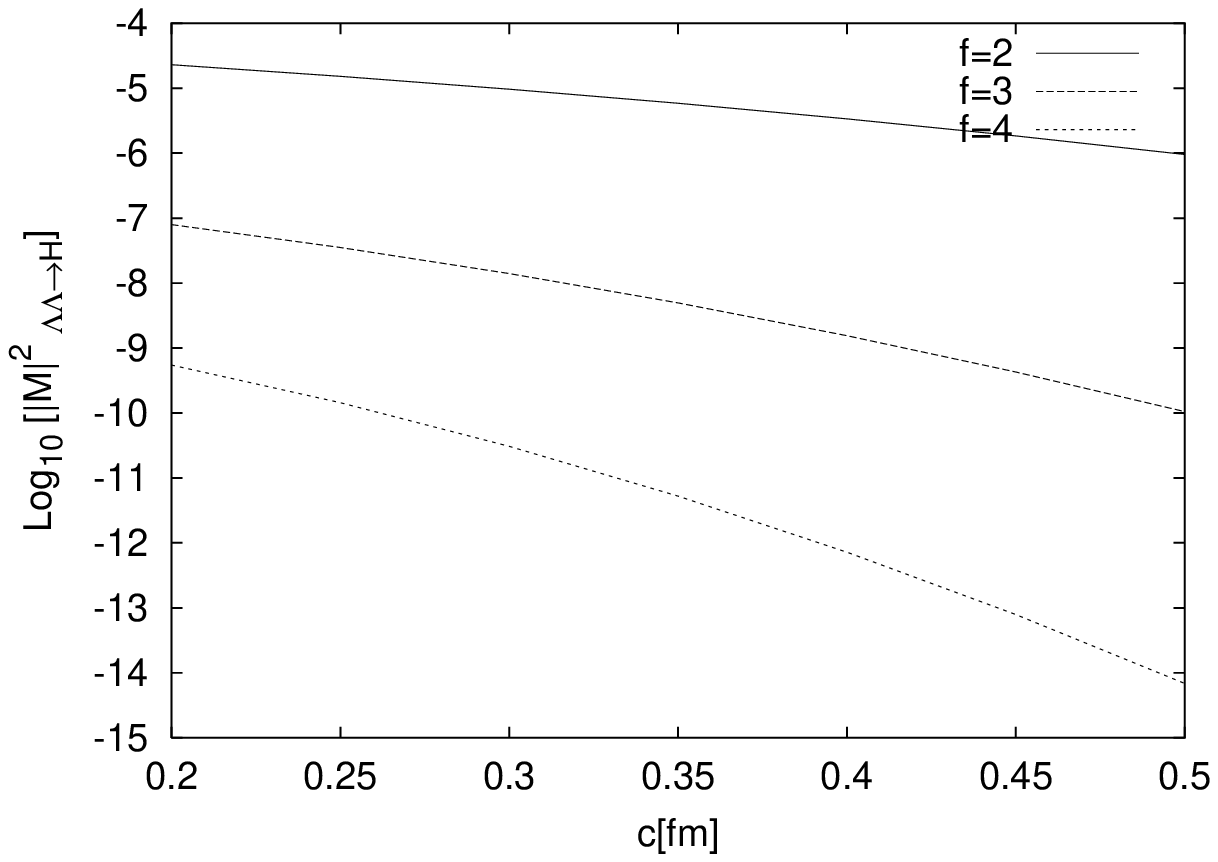,width=8cm}}{\footnotesize\textbf{Figure
2:} Log$_{10}$ of $|{\cal M}|^2_{\Lambda \Lambda \rightarrow H}$
versus hard core radius in fm, for f=2, 3, 4.}
\vspace{5mm}\\

\subsection{Calculation of the $|{\cal M}|_{BB\rightarrow \Lambda\Lambda X}$ matrix
element}

Transition of a two nucleon system to $\Lambda \Lambda$ requires
two strangeness changing weak reactions. Possible $\Delta S=1$
sub-processes to consider are a weak transition with emission of
a pion or lepton pair and an internal weak transition.  These
are illustrated in Fig. 3 for a three quark system. We estimate
the amplitude for each of the sub-processes and calculate the
overall matrix element for transition to the $\Lambda \Lambda$
system as a product of the sub-process amplitudes. \\

\vspace{5mm}
\centerline{\epsfig{file=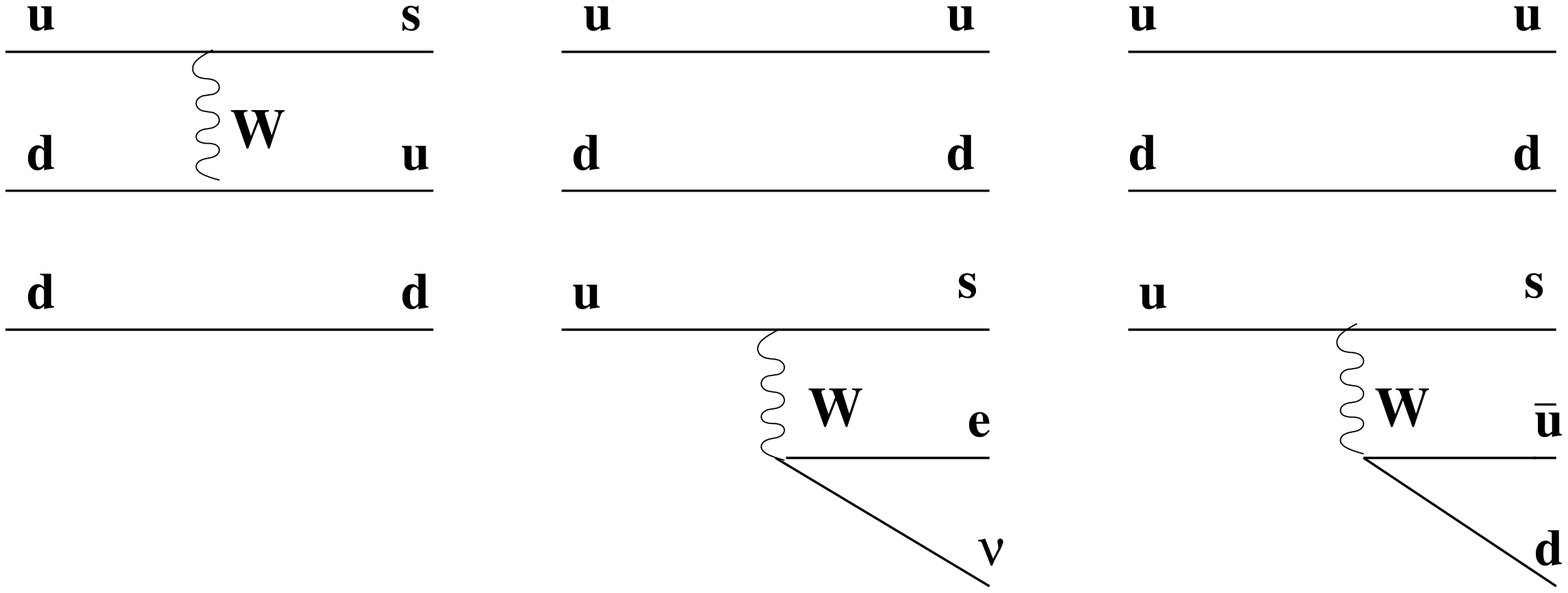,width=6cm}}{\footnotesize\textbf{Figure
3:} Some relevant weak transitions for $NN \rightarrow HX$ }
\vspace{5mm}\\

The matrix element for weak pion emission is estimated from the
$\Lambda\rightarrow N \pi$ rate: \beq |{\cal
M}|^2_{\Lambda\rightarrow N \pi}=\frac {1}{(2\pi )^4} ~ \frac
{2m_{\Lambda} }{\Phi _2} \frac {1}{\tau _{\Lambda\rightarrow N
\pi}}=0.8 \times 10^{-12} \quad {\rm GeV}^{2}. \eeq By crossing
symmetry this is equal to the desired $|{\cal
M}|^2_{N\rightarrow \Lambda \pi}$, in the approximation of
momentum-independence which should be valid for the small
momenta in this application. Analogously, for lepton pair
emission we have \beq |{\cal M}|^2_{\Lambda\rightarrow N e\nu }=
\frac {1}{(2\pi )^4}~\frac {2 m_{\Lambda}  } {\Phi _3 }\frac
{1}{ \tau _{\Lambda\rightarrow N e\nu }} =3.0 \times 10^{-12}.
\eeq

The matrix element for internal conversion, $(uds) \rightarrow
(udd)$, is proportional to the spatial nucleon wave function
when two quarks are at the same point: \beq |{\cal
M}|_{\Lambda\rightarrow N} \approx <\psi _{\Lambda }|\delta^3
(\vec {r}_1-\vec {r}_2)|\psi _N > \frac {G_F \sin \theta _c \cos
\theta _c}{m_q}, \eeq where $m_q$ is the quark mass introduced
in order to make the 4 point vertex amplitude
dimensionless\cite{yaouanc}. The expectation value of the delta
function can be calculated in the harmonic oscillator model to
be \beq \label{delta1} <\psi _{\Lambda }|\delta^3 (\vec
{r}_1-\vec {r}_2)|\psi _N >~ = \left(\frac {\alpha _B}{\sqrt {2
\pi }}\right)^3=0.4 \times 10^{-2} ~~ {\rm GeV}^3. \eeq The
delta function term can be also inferred phenomenologically in
the following way, as suggested in \cite{yaouanc}. The Fermi
spin-spin interaction has a contact character depending on
$~\vec {\sigma _1}\vec { \sigma_2}/m^2 _q \delta(\vec {r}_1-\vec
{r}_2)$, and therefore the delta function matrix element can be
determined in terms of electromagnetic or strong hyperfine
splitting:
\begin{eqnarray}
(m_{\Sigma ^0}-m_{\Sigma ^+} )-(m_n-m_p)=\alpha \frac {2\pi
}{3m^2 _q}<\delta^3(\vec {r}_1-\vec {r}_2)>\\m_{\Delta} -m_N=
\alpha _S \frac {8\pi }{3 m^2 _q} <\delta^3(\vec {r}_1-\vec
{r}_2)>.
\end{eqnarray}
where $m_q=$ taken to be $m_N/3$ is the quark mass. Using the
first form to avoid the issue of scale dependence of $\alpha_S$
leads to a value three times larger than predicted by the method
used in eqn \ref{delta1}, namely: \beq \label{delta2} <\psi
_{\Lambda }|\delta^3 (\vec {r}_1-\vec {r}_2)|\psi _N> ~ =
1.2\times 10^{-2} \quad {\rm GeV}^3. \eeq We average of the
expectation values (\ref{delta1}) and (\ref{delta2}) and find
\beq |{\cal M}|^2_{\Lambda\rightarrow N}=4.4 \times 10^{-15}.
\eeq In this way we have roughly estimated all the matrix
elements for the relevant sub-processes based on
weak-interaction phenomenology.

%\vspace{5mm}
%\centerline{\epsfig{file=NNH.eps,width=6cm}}{\footnotesize\textbf{Figure
%2:} Dominant channels: $NN \rightarrow H\pi $ and  $NN
%\rightarrow H\pi \pi$ }
%\vspace{5mm}\\

\subsection{Nuclear decay rates} \label{nuclifetime}

NN $\rightarrow$ HX requires two weak reactions. For the process
$A_{NN}\rightarrow A'_{H}\pi \pi$, the  rate is thus
approximately
\begin{eqnarray}
\Gamma_{A_{NN} \rightarrow  A'_{H} \pi \pi }&\approx &K^2 \frac
{(2\pi )^4} {2 (2m_{N}) }~ \Phi_3\\ \nonumber 
&\times & \left(
\frac { |{\cal M}|_{N\rightarrow \Lambda \pi} ^2 |{\cal
M}|_{\Lambda \Lambda \rightarrow H} } {(2m_{\Lambda }-m_H )^2
}\right) ^2
\end{eqnarray}
where the denominator is introduced in the spirit of the
$\Lambda \Lambda$ pole approximation, to make the 4 point vertex
amplitude dimensionless. The lifetime for this decay is \beq
\tau_{A_{NN}\rightarrow A'_{H}\pi \pi}\approx \frac {0.03}{K^2
|{\cal M}|^2_{\Lambda \Lambda \rightarrow H}} \quad {\rm yr},
\eeq taking $m_H = 1.5$ GeV in the phase space factor.  For the
process with one pion emission and an internal conversion, the
rate estimate is
\begin{eqnarray}
\Gamma_{A_{NN}\rightarrow A'_{H}\pi}&\approx &K^2\frac {(2\pi
)^4}{2 (2m_{N})}~\Phi_2 \\ \nonumber 
&\times &(|{\cal
M}|_{N\rightarrow \Lambda \pi} |{\cal M}|_{N\rightarrow \Lambda}
|{\cal M}|_{\Lambda \Lambda \rightarrow H})^2
\end{eqnarray}
leading to the lifetime for $m_H = 1.5$ GeV of
\beq
\tau_{A_{NN}\rightarrow A'_{H} \pi}\approx \frac {2 \times
10^{-3}}{K^2 |{\cal M}|^2_{\Lambda \Lambda \rightarrow H}} \quad
{\rm yr}. \eeq

If $m_H \gsi 1740$ MeV, pion emission in a nucleus is
kinematically suppressed and the relevant final states are $e^+
\nu$ or $\gamma$; we now calculate these rates, taking $m_H =
1.8$ GeV. For the transition $A_{NN}\rightarrow A' _H e\nu$, the
rate is
\begin{eqnarray}
\Gamma_{A_{NN} \rightarrow  A'_{H}e\nu }&\approx &K^2\frac {(2\pi
)^4}{2 (2m_{N})}\Phi_3 \\ \nonumber 
&\times &(|{\cal
M}|_{N\rightarrow \Lambda e\nu} |{\cal M}|_{N\rightarrow
\Lambda} |{\cal M}|_{\Lambda \Lambda \rightarrow H})^2.
\end{eqnarray}
In this case, the nuclear lifetime is \beq \label{enu}
\tau_{A_{NN}\rightarrow A'_{H} e\nu}\approx \frac {70}{K^2
|{\cal M}|^2_{\Lambda \Lambda \rightarrow H}} \quad {\rm yr}.
\eeq For $A_{NN}\rightarrow A' _H \gamma$, the rate is
approximated as
\begin{eqnarray}
\Gamma_{A_{NN}\rightarrow A'_{H}\gamma }&\approx &K^2 (2\pi )^4
\frac {\alpha _{EM} m^2 _q}{2 (2m_{N})} \\ \nonumber &\times &
\Phi_2(|{\cal M}|^2 _{N\rightarrow \Lambda} |{\cal M}|_{\Lambda
\Lambda \rightarrow H})^2.
\end{eqnarray}
leading to the lifetime 
\beq 
\tau_{A_{NN}\rightarrow A'_{H}
\gamma}\approx \frac {2~10^3}{K^2 |{\cal M}|^2_{\Lambda \Lambda
\rightarrow H}} \quad {\rm yr}. 
\eeq 
One sees from Fig. 1 that a
lifetime bound of $\gsi {\rm few}~10^{29}$ yr is not a very
stringent constraint on this scenario if $m_H$ is large enough
that pion final states are not allowed.  E.g., with $K^2 =
1/1440$ the rhs of eqn (\ref{enu}) is $\gsi {\rm few}~10^{29}$
yr, for the a hard core radius of 0.45 fm and $r_H \approx 1/5
~r_N$ -- in the middle of the range expected based on the
glueball analogy.  If $m_H$ is light enough to permit pion
production, experimental constraints are much more powerful.
$m_H \lsi 1740$ MeV is disfavored but not excluded; the allowed
region in the $f$-hard core radius plane may be reasonable,
depending on how strong limits SuperK can give.

\begin{table}[hpb]
\small
%\tiny
\caption{The final particles and momenta for nucleon-nucleon
transitions to H in nuclei. For the 3-body final states marked
with *, the momentum given is for the configuration with H
produced at rest.} \label{t1}
\begin{center}
\begin{tabular}{|c||c|c||c|}
\hline

mass        & final state & final momenta  & partial lifetime  \\
$m_H$ [GeV] & A' + H +    & p [MeV]        & $ \times K^2|{\cal
M}|^2 _{\Lambda \Lambda \rightarrow H}$ [yr] \\ \hline

1.5         &  $\pi $     & 318            & $2~10^{-3}$    \\ \hline
1.5         &  $\pi \pi$  & 170*           & 0.03           \\ \hline
1.8         &  $e \nu$    & 48*            & 70             \\ \hline
1.8         &  $\gamma$   & 96             & $2~10^3$       \\ \hline \hline

\end{tabular}
\end{center}
\end{table}

\section{Conclusions} \label{summary}

We have considered the stability of nuclei and hypernuclei with
respect to conversion to an H dibaryon.  If the binding of the H
dibaryon is strong, possibly resulting in $m_H < 2 m_N$ as
conjectured in refs. \cite{f:stableH,kf:Lam1405}, then the size
of the H is expected to be much smaller than the size of a
nucleon and comparable to the size of a glueball: $r_H \approx
r_G \approx (1/6 - 1/4)~r_N$.  We used the Isgur-Karl
wavefunctions for quarks in baryons and the H, and the
Bethe-Goldstone wavefunction for nucleons in a nucleus, to
obtain a rough estimate of the wavefunction overlap for the
process $A_{BB} \rightarrow A'_{H} X$.  We find that observation
of $\Lambda$ decays in double-$\Lambda$ hypernuclei does not
exclude an H -- stable or not -- as long as $r_H \lsi 1/3~ r_N$.

Combining our wavefunction overlap estimates with
phenomenological weak interaction matrix elements, permits the
lifetime for conversion of nuclei to H to be estimated.  These
estimates have uncertainties of greater than an order of
magnitude: the weak interaction matrix elements are uncertain to
a factor of a few, factors of order 1 were ignored, a crude
statistical estimate for the flavor-spin overlap was used, mass
scales were set to $m_N/3$, and most importantly, the
calculation of the wavefunction overlap used models which surely
oversimplify the physics. While the overlap is highly uncertain
because it depends on nuclear wavefunctions and hadronic
dynamics which are not adequately understood at present, the
enormous suppression of H production which we found in this
calculation forces us to conclude that an absolutely stable H is
not excluded by these considerations.

SuperK can place important constraints on the conjecture of an
absolutely stable H, or conceivably discover evidence of its
existence, through observation of the pion(s), positron, or
photon produced when two nucleons in an oxygen nucleus convert
to an H. We estimated that SuperK could achieve a lifetime limit
$\tau \gsi {\rm few} ~10^{29}$ yr. Until the properties of the H
and the dynamics of production of the H in nuclei are better
understood, this limit would be insufficient to rule out a
stable H. However such a sensitivity would access the estimated
lifetime range for $m_H \gsi 1740$ MeV and $r_H \approx 1/5
~r_N$, and an experimental search is warranted.

The research of GRF was supported in part by NSF-PHY-0101738.
GRF acknowledges helpful conversations with many colleagues,
particularly G. Baym, A. Bondar, T. Kajita, M. May, M.
Ramsey-Musolf, and P. Vogel.  GZ wishes to thank Marko Kolanovic for useful advice and is grateful to
Emiliano Sefusatti for many helpful comments.

\bibliographystyle{unsrt}
\bibliography{stabilityH}

\end{document}